\documentclass{osa-article}
\usepackage{bbold}
\usepackage[normalem]{ulem}

\journal{oe}

\articletype{Research Article}

\begin{document}

\title{Go-and-return phase encoded SR QKD and its security consideration}

\author{Vadim Rodimin,\authormark{1, 2, 3*} Andrey Tayduganov,\authormark{1, 2} Dmitry Kronberg,\authormark{3,4,5} Yury Durkin,\authormark{1,2} Alexey Zharinov,\authormark{1,2} and Yury Kurochkin\authormark{1,2,3}}

\address{
\authormark{1}QRate, Skolkovo, Moscow 143025, Russia \\
\authormark{2}NTI Center for Quantum Communications, National University of Science and Technology MISiS, Leninsky prospekt 4, Moscow 119049, Russia \\
\authormark{3}Russian Quantum Center, Skolkovo, Moscow 143025, Russia \\
\authormark{4}Steklov Mathematical Institute of Russian Academy of Sciences, Moscow 119991, Russia \\
\authormark{5}Institute of Physics and Technology (National Research University), Institutskii per. 9, 141701 Dolgoprudnyi, Moscow region, Russia
}
\email{\authormark{*}v.rodimin@goqrate.com} 



\begin{abstract}

In this work we study the security of coherent-state quantum key distribution with a strong reference pulse. The consideration is based on a powerful soft filtering attack and uses realistic parameters of the equipment. Our model allows us to relate the secure key generation rate and distance with the energy of signal and reference pulses. We propose a two-pass quantum key distribution phase encoding optical scheme with a strong reference pulse. This simple for realization scheme is experimentally tested showing stable operation. Here we describe the tuning technique and solution for the problem of weak signal pulse shadowing by a bright reference pulse with the relative intensity difference of more than 60\,dB between them.
\end{abstract}


\section{Introduction}

Quantum key distribution (QKD) is a technology for shared random secret key generation between two distant parties, usually called Alice (transmitter) and Bob (receiver). The security of QKD is based on the fundamental laws of quantum physics what makes the encryption that uses QKD theoretically absolutely secure. Due to essential breakthroughs in high-speed signal processing and miniaturization of photoelectric devices in recent decades, QKD technology significantly advanced. Provably secure commercial QKD systems are already available on the market. The next breakthrough can be expected with the development and cheapening of single-photon sources, and with the appearance of a commercially available high-fidelity quantum memory.

So far, in commercial QKD systems the weak attenuated laser pulses are used instead of single photons.
The presence of multi-photon states in a laser pulse and losses in quantum channel (QC) allow the eavesdropper (Eve) to carry out a photon number splitting (PNS) attack \cite{Brassard00} and obtain the full (or at least significant part of) information without being detected. Eve performs a non-demolition photon number measurement, blocks pulses containing one photon, and splits all multi-photon pulses, leaving a part in her quantum memory for subsequent information retrieval after basis reconciliation and delivering the rest to Bob via {\it lossless} channel. If it is required to provide Bob the expected raw key rate, she can adjust the fraction of blocked single-photon pulses according to real optical scheme parameters. This QKD attack strategy is so efficient that the secure key generation via lossy optical fiber using weak attenuated laser pulses is limited to just a few dozens of kilometers.

Various finesses have been developed to counteract the PNS attack. The most popular and efficient solution is the decoy state method \cite{Hwang03,Wang05,Lo05,Ma05}, which  significantly increases the QKD distance and has already become a non-official standard. However, there is another way to avoid the problem as was proposed earlier in the original B92 protocol \cite{B92} -- to send also a bright reference pulse which must be {\it always} detected on Bob's side. Although inferior to the decoy state method in terms of maximal distance, it wins in terms of complexity of key post-processing and technical implementation. Typically, the information bit in such scheme is encoded in the relative phase between the weak signal pulse (SP) and the strong reference pulse (SRP). In practice, the information-carrying dim SP is mixed with high-intensity SRP, forming a common effective quantum state. Using SRP makes it impossible for Eve to perform the PNS attack, since being projected onto the Fock basis SP looses its phase relation with SRP. In this case, Eve loses the ability to block any pulse, since SRP must always reach Bob.

Nevertheless, Eve still can extract some partial information from multi-photon pulses. In Ref.~\cite{Acin04} the authors studied various protocols, including B92 and BB84 with SRP (which is equivalent to the 4+2 protocol \cite{Huttner95}), resistant against the PNS attack. By taking the limit of infinite SRP intensity for very large distances, the authors made a rough estimation of Eve's information and found it to be negligible, thus concluding the security of the B92 protocol against PNS attacks. It was explicitly stressed out that the SRP mean photon number on Bob's side has to be significant, and that it is important for Alice to increase the SRP intensity with distance.

Although with all these arguments the presence of SRP seems to make the considered protocols secure against the PNS attack, one requires a rigorous mathematical proof for all types of attacks, deriving a corresponding explicit secret key rate formula.
The first unconditional security proof of the B92 protocol with perfect single-photon source was presented by Tamaki {\it et al}. in Ref.~\cite{Tamaki03} under assumption that Alice's and Bob’s devices are ideal and the quantum channel is lossless. Following the idea in Ref.~\cite{Tamaki03}, the security was later proved for lossy and noisy channel but with assumption that Bob has a perfect detector that discriminates between single-photon and vacuum or multi-photon states \cite{Tamaki04}.
The B92 version with SRP was considered in Ref.~\cite{Koashi04}, where the authors made a modification of Bob's optical scheme by introducing a second local oscillator, locked to the SRP mode, in order to simplify the analysis. Avoiding this modification due to its practical implementation complexity, another proof was made in Ref.~\cite{Tamaki09} in which Bob is supposed to have two types of detectors: one type can tell whether the photon number is in a certain interval, and the other type can discriminate among vacuum, single-photon and multi-photon events. Since that time, according to our knowledge, there were no new and more general unconditional security studies that could be applied to existing practical realizations of B92 or BB84 with SRP. 

To the best of our knowledge, SR QKD systems have not gone beyond laboratory demonstrations of capabilities, with the exception of the subcarrier system \cite{Gleim2016}, which has been brought to the level of practical applications. This setup uses frequency bit coding \cite{Bloch2007} varying the sidebands intensity in different measurement bases and the central frequency carrier plays the role of SRP. The authors introduce ``low intensity reference'' and point out that a hundred photons in SRP are enough to withstand the PNS attack. The absence in the literature of a unified view on SRP energy prompted us to find an explicit distance dependence on photon numbers in SRP for secure key generation.

This work is organized as follows. In Sec.~\ref{sec:attack} we consider the soft filtering attack instead of PNS, estimate intensity of the SRP for secure key distribution and the secret key rate that can be achieved. In Sec.~\ref{sec:experiment} present our optical scheme and experimental results. In Sec.~\ref{sec:discussion} we mention some unsolved experimental difficulties and discuss the optimal parameter choice comparing the SRP-based QKD protocols. We summarize the results of our work and conclude in Sec.~\ref{sec:conclusion}.

\section{Secret key rate estimation}\label{sec:attack}

\subsection{Soft filtering attack}

Due to the lack of an unconditionally secure key rate formula for the real setup without any special source/detector assumption, in order to estimate optimal weak signal and strong reference pulse intensities, in this work we consider Eve's attack in the form of quantum soft filtering operation \cite{Kronberg18,Kronberg19}, which is a more general version of filtering (see e.g. Ref.~\cite{Acin04}). This unitary operation provides probabilistic information extraction and is an interpolation between the standard beam-splitting (BS) attack and unambiguous state discrimination (USD) technique.
We consider a scenario when Eve does not have access to Bob's device and hence cannot control his detectors or replace them by ideal ones.
Now let us describe the steps of attack.
 \begin{enumerate}
 	\item Eve intercepts the coherent state $|A\rangle\otimes|\pm\alpha\rangle$, consisted of SRP ($A$) and SP ($\pm\alpha$), attaches her ancilla and performs a two-outcome soft filtering on $|\Psi_{0,1}\rangle=|A\rangle\otimes|\pm\alpha\rangle\otimes|{\rm ancilla}\rangle$, denoted by a unitary transformation $U$:
		\begin{equation}
			|\Psi_{0,1}^{\rm filter}\rangle = U |\Psi_{0,1} \rangle = \sqrt{p}\, |A_s\rangle \otimes |\pm\alpha_s\rangle \otimes |s\rangle + \sqrt{1-p}\, |A_f\rangle \otimes |\pm\alpha_f\rangle \otimes |f\rangle \,,
			\label{eq:filtering}
		\end{equation}
		obtaining either {\it success} $|s\rangle$ or {\it fail} $|f\rangle$ outcomes ($\langle s|f\rangle=0$) in her auxiliary system with probabilities $p$ and $1-p$ respectively. Here {\it success}({\it fail}) denotes obtaining more(less) distinguishable non-orthogonal signal states $|\pm\alpha_{s(f)}\rangle$ compared to the initial $|\pm\alpha\rangle$.
		The corresponding intensities of the states before and after filtering are $|A|^2=\nu$, $|\alpha|^2=\mu$, $|\alpha_s|^2=a\mu$ and $|\alpha_f|^2=b\mu$ with $a>1$, $b<1$.
		
		From the unitarity condition $\langle\Psi_0^{\rm filter}|\Psi_1^{\rm filter}\rangle=\langle\Psi_0|U^\dagger U|\Psi_1\rangle=\langle\Psi_0|\Psi_1\rangle$, simplified to
		\begin{equation}
	        p \langle\alpha_s|-\alpha_s\rangle + (1 - p) \langle\alpha_f|-\alpha_f\rangle = \langle\alpha|-\alpha\rangle \,,
		\end{equation}
		one obtains the following constraint on Eve's attack parameter space, namely the success probability $p$ and the amplification/attenuation coefficients $a/b$:
        \begin{equation}
            p\, e^{-2a\mu} + (1 - p)\, e^{-2b\mu} = e^{-2\mu} \,.
	        \label{eq:unitarity_cond}
        \end{equation}
		
	\item In case of {\it success}({\it fail}) Eve splits the outcome state and sends $|B_{s(f)}\rangle\otimes|\pm\beta_{s(f)}\rangle$ to Bob via lossless channel, storing the remaining part in her quantum memory for future collective measurement. Apparently, the intensities of sent states have to be $|B_{s(f)}|<|A_{s(f)}|$, $|\beta_{s(f)}|<|\alpha_{s(f)}|$. To remain undetected it is very important for Eve to take into account all points of Bob's measurement procedure, listed below.

		\begin{enumerate}
			\item Bob delays the signal pulse and makes it interfere with a fraction $\mu/\nu$ of the strong pulse. Therefore, in order not to spoil the interference, Eve must always set
			    \begin{equation}
			        {|\beta_{s(f)}|^2 \over |B_{s(f)}|^2}= {\mu \over \nu} \,.
			    \end{equation}
			    Otherwise she will increase Bob's error rate and hence will be detected.

			\item Since Bob monitors the intensity of detected SRP, Eve has to adjust $B_{s,f}$ such that $|B_{s,f}|^2\in[\nu_{\min}^\prime,\nu_{\max}^\prime]$ with $\nu_{\rm max(min)}^\prime=\nu^\prime(1\pm\delta)$, where $\nu^\prime=\nu\times10^{-0.2L/10}$ is the expected SRP intensity on receiver's side, and $\delta$ is determined by precision of Bob's monitoring device.
			
			\item From both conditions above follows $|\beta_{s,f}|^2\in[\mu_{\min}^\prime,\mu_{\max}^\prime]$ with similarly defined limits $\mu_{\rm max(min)}^\prime=\mu^\prime(1\pm\delta)$ and expected signal intensity $\mu^\prime=\mu\times10^{-0.2L/10}$.
			
			\item Eve has to provide the expected raw key rate, determined by Bob's USD measurement (for the rate estimation in protocols based on non-orthogonal states see Appendix~\ref{app:B92}). This gives the following constraint on sent intensities $|\beta_{s,f}|^2$:
			    \begin{equation}
	                p \big( 1 - e^{-2\eta|\beta_s|^2} \big) + (1 - p) \big( 1 - e^{-2\eta|\beta_f|^2} \big) = 1 - e^{-2\eta\mu^\prime} \,,
	                \label{eq:key_rate_cond}
                \end{equation}
                where $\eta$ is the Bob's detector efficiency.
		\end{enumerate}
 \end{enumerate}
 Note that due to the constraints \eqref{eq:unitarity_cond} and \eqref{eq:key_rate_cond} the initial five-dimensional parameter space $\{p,a,b,|\beta_s|,|\beta_f|\}$ is reduced to a three-dimensional one, e.g. $\{b,|\beta_s|,|\beta_f|\}$.

In principle, if Eve gets more(less) distinguishable states $|\pm\alpha_{s(f)}\rangle$ after filtering, it is natural for her to send Bob as stronger(weaker) pulse as possible in order to increase(decrease) his chance of having a conclusive measurement result. Therefore, in this work we assume for simplicity $|\beta_{s(f)}|^2=\mu_{\max(\min)}^\prime$. In this way we are left with only one free parameter $b$. Study of a more general and optimal attack scenario we postpone for future publication.

To summarize the described attack, we find the maximum amount of information that Eve can extract from her states $|\pm\epsilon_{s(f)}\rangle$ of intensity $|\epsilon_{s(f)}|^2=|\alpha_{s(f)}|^2-|\beta_{s(f)}|^2$ \cite{Kronberg18}:
\begin{equation}
    \begin{split}
	I_E &= \max_b \bigg\{
		{p \big(1 - e^{-2\eta|\beta_s|^2}\big) \chi(|\epsilon_s|^2)
		+ (1 - p) \big(1 - e^{-2\eta|\beta_f|^2}\big) \chi(|\epsilon_f|^2)
		\over p \big(1 - e^{-2\eta|\beta_s|^2}\big) + (1 - p) \big(1 - e^{-2\eta|\beta_f|^2}\big)}
		\bigg\} \\
		&= \max_b \bigg\{
		{p \big(1 - e^{-2\eta\mu_{\max}^\prime}\big) \chi(a\mu - \mu_{\max}^\prime)
		+ (1 - p) \big(1 - e^{-2\eta\mu_{\min}^\prime}\big) \chi(b\mu - \mu_{\min}^\prime)
		\over 1 - e^{-2\eta\mu^\prime}}
		\bigg\} \,.
	\end{split}
	\label{eq:I_E_final}
\end{equation}
where $p$ and $a$ are determined from Eqs.~\eqref{eq:key_rate_cond} and \eqref{eq:unitarity_cond} respectively and are given in Appendix~\ref{app:filtering_constraints}; Holevo $\chi$--value and Shannon binary entropy $H$ are defined as
\begin{equation}
	\chi(\mu) = H \bigg( {1 - e^{-2\mu} \over 2} \bigg) \,, \quad H(x) = -x \log_2 x - (1 - x) \log_2 (1 - x) \,.
\end{equation}
For fixed setup configuration ($L$, $\mu$, $\delta$) Eve adjusts the attenuation parameter $b$, providing her maximum possible information.

\subsection{Protocol parameter optimization}

The secret key rate is determined by
\begin{equation}
	R_{\rm sec} = R_{\rm raw} (I_{A:B} - I_E) \simeq \xi f (1 - e^{-2\eta\mu^\prime}) \big[ 1 - f_{\rm ec} H({\rm QBER}) - I_E \big] \,,
	\label{eq:R_sec}
\end{equation}
where $\xi=1$ for B92 and $\xi=1/2$ for BB84 schemes, $f$ is the pulse repetition frequency, $f_{\rm ec}\simeq1.2$ is the typical error correction efficiency.

Assuming for simplicity that Eve's contribution to Bob's quantum bit error rate (QBER) is negligible, it can be estimated as follows,
\begin{equation}
	{\rm QBER} = {p_{\rm dc} + p_{\rm opt} (1 - e^{-2\eta\mu^\prime}) \over 2p_{\rm dc} + 1 - e^{-2\eta\mu^\prime}} \,,
	\label{eq:QBER}
\end{equation}
with detector efficiency $\eta=20\%$, dark count (per detector) probability $p_{\rm dc}=2\times10^{-5}$ and optical error probability $p_{\rm opt}=2\%$.

The uncertainty of the SRP intensity measurement can be estimated by noise-equivalent power (NEP) of Bob's monitoring detector as
\begin{equation}
	\Delta\nu^\prime \simeq {{\rm NEP} \sqrt{\tau} \over E_{\rm photon}} = {{\rm NEP} \sqrt{\tau} \lambda \over hc} \,,
\end{equation}
with pulse width $\tau=5$\,ns, NEP=25\,pW$\cdot$Hz$^{-1/2}$ and $\lambda=1550$\,nm. Defining the relative attenuation $t$ between strong and weak pulses as $\mu/\nu=10^{-t/10}\ll1$, one obtains
\begin{equation}
	\delta = {\Delta\nu^\prime \over \nu^\prime} \simeq {{\rm NEP} \sqrt{\tau} \lambda \over \mu^\prime hc} \times 10^{-t/10} \simeq {1.38 \over \mu} \times 10^{4 + 0.02L - 0.1t} \,.
	\label{eq:delta}
\end{equation}
Note that $\delta$ depends on $\mu$ with fixed $t$.

\begin{figure}[t!]\centering
	\includegraphics[width=0.45\textwidth]{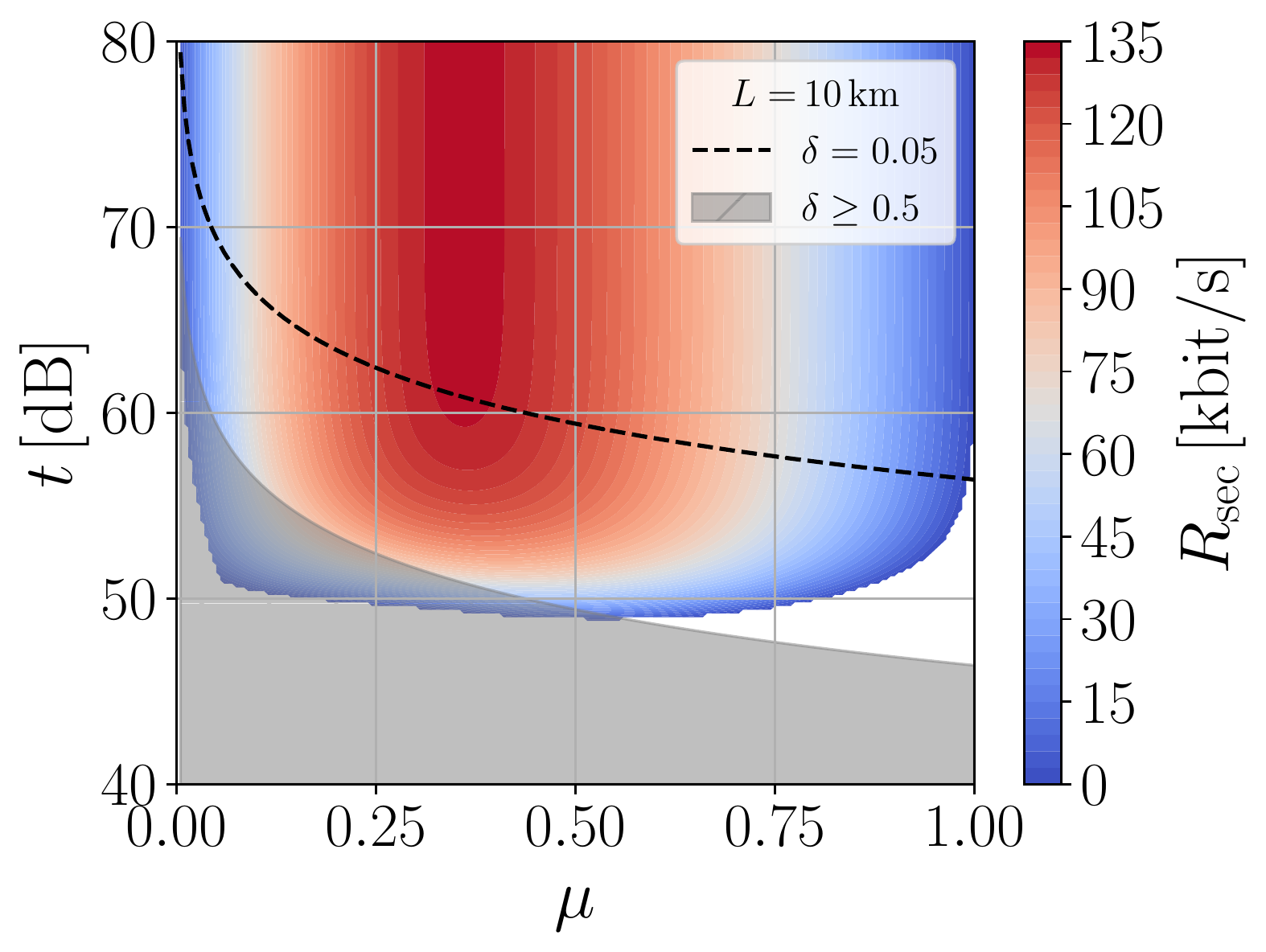} \hspace{3mm}
	\includegraphics[width=0.43\textwidth]{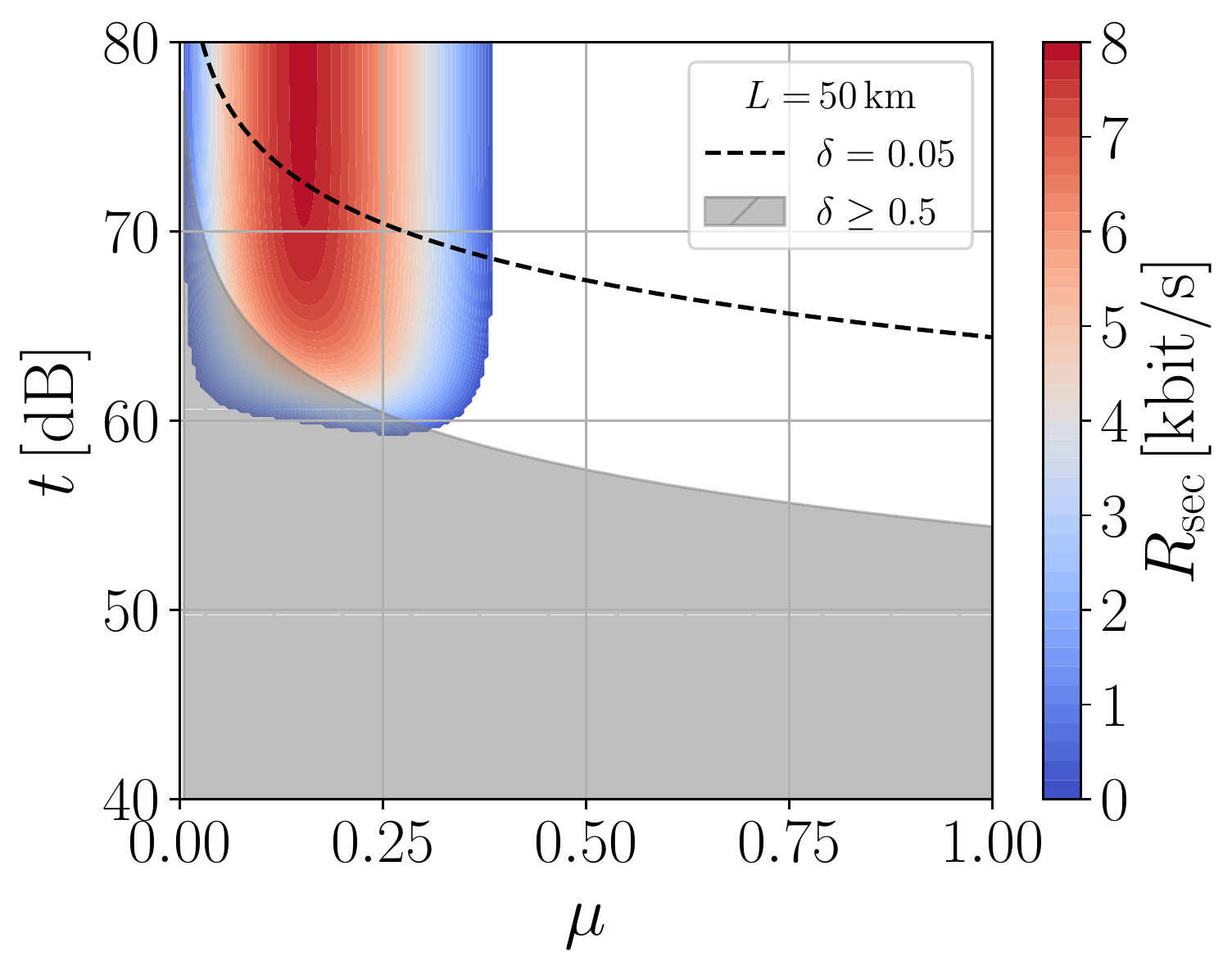}
	\caption{Contour plot of secure key generation rate as function of signal pulse intensity $\mu$ and its attenuation $t$ with respect to SRP for optic line lengths 10\,km (left) and 50\,km (right). Grey region represents the parameter space for which the SRP monitoring precision $\delta$, determined by Eq.~\eqref{eq:delta}, is worse than 50\% -- such measurement can be considered as unacceptable.}
	\label{fig:R_sec_contour}
\end{figure}

\begin{figure}[t!]\centering
	\includegraphics[width=0.45\textwidth]{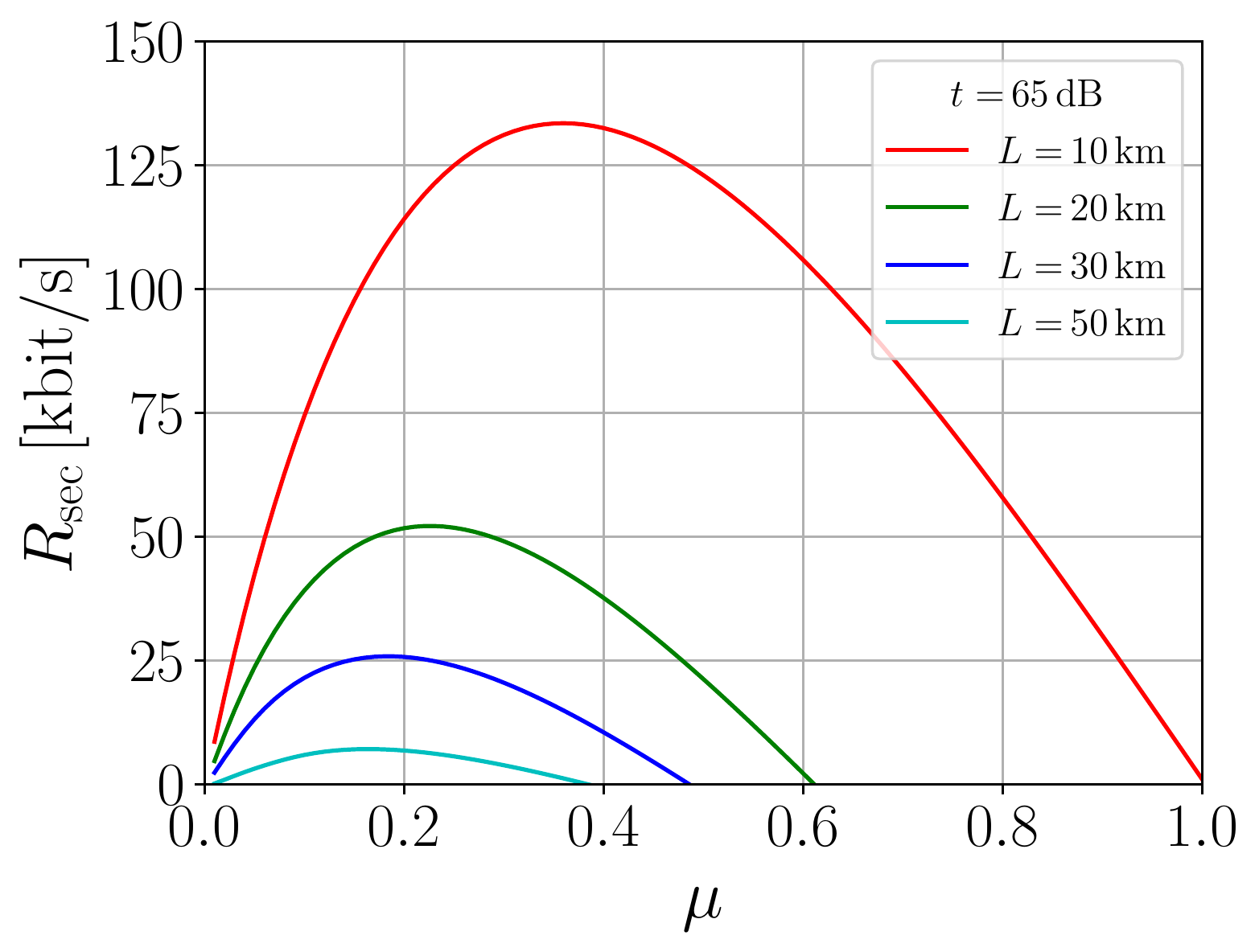} \hspace{3mm}
	\includegraphics[width=0.45\textwidth]{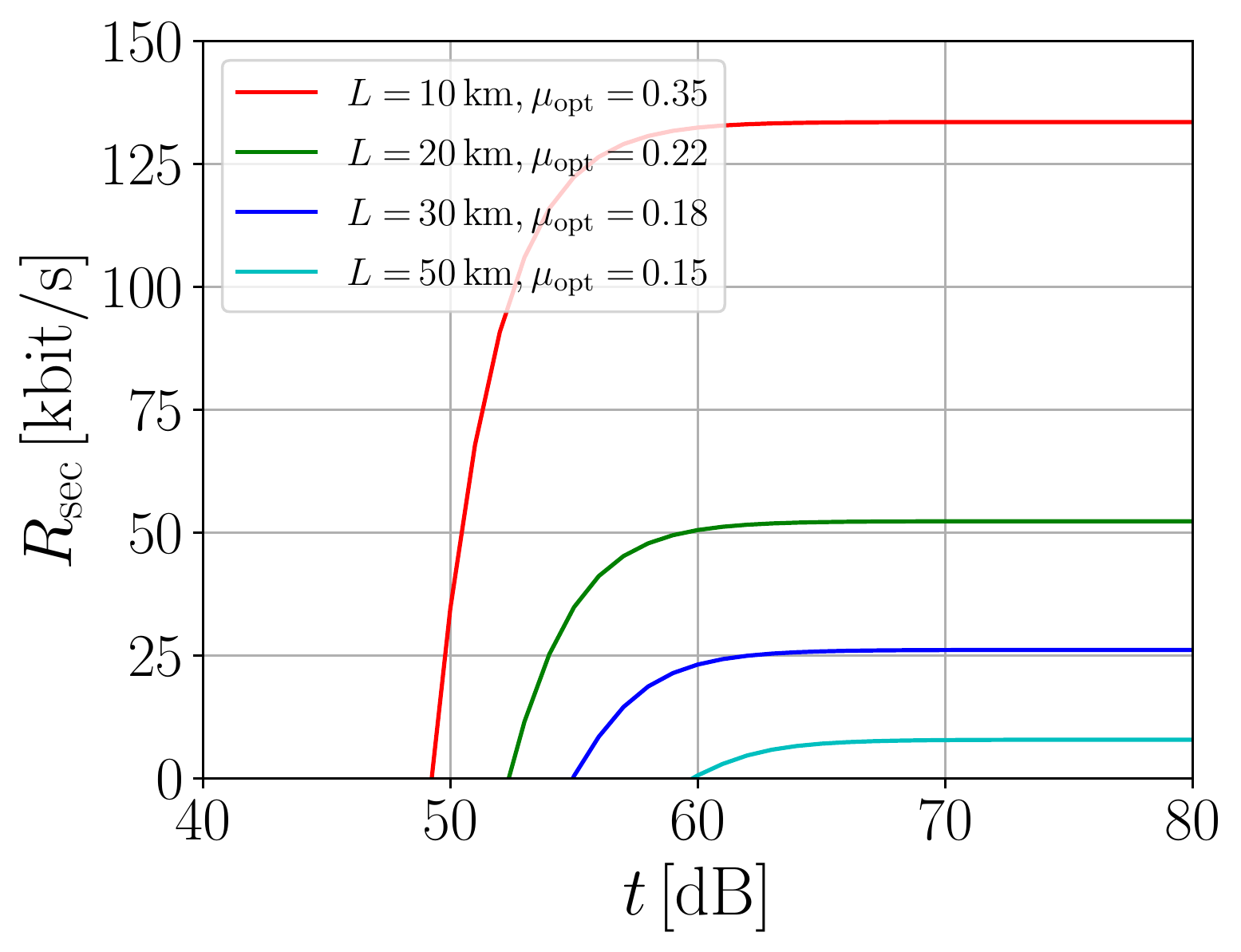}
	\caption{Secure key generation rate as function of signal pulse intensity $\mu$ (left) and attenuation $t$ with respect to SRP (right). The optimal intensities $\mu_{\rm opt}$ are determined from maximums of $R_{\rm sec}(\mu)$ in the left plot.}
	\label{fig:R_sec-mu-t}
\end{figure}

In order to determine optimal parameters for our experimental QKD setup, we present in Fig.~\ref{fig:R_sec_contour} two contour plots of $R_{\rm sec}$ as function of signal intensity $\mu$ and attenuation $t$ for $L=10,50$\,km. At each $(\mu,t)$--point Eve's information, determined by Eq.~\eqref{eq:I_E_final}, is maximized by numerical search for optimal $b$--value. One can see from the plots that for Alice and Bob the optimal intensity $\mu_{\rm opt}$ that provides them maximum rate with SRP monitoring precision $\delta\lesssim5\%$ is around 0.3(0.1), and $t$ starts from about 60(70)\,dB for 10(50)\,km. For more accurate determination of $\mu_{\rm opt}$ see the plot on the left in Fig.~\ref{fig:R_sec-mu-t}. The plot on the right demonstrates that starting from some value of attenuation $R_{\rm sec}$ becomes constant, and consequently further increase of $t$ does not improve the final key rate. Therefore, in our experimental setup we set $t=65$\,dB as universal for $L\lesssim50$\,km.

\section{Experimental realization}\label{sec:experiment}

\subsection{Experimental setup}

\begin{figure}[ht!]
\centering\includegraphics[width=0.99\textwidth]{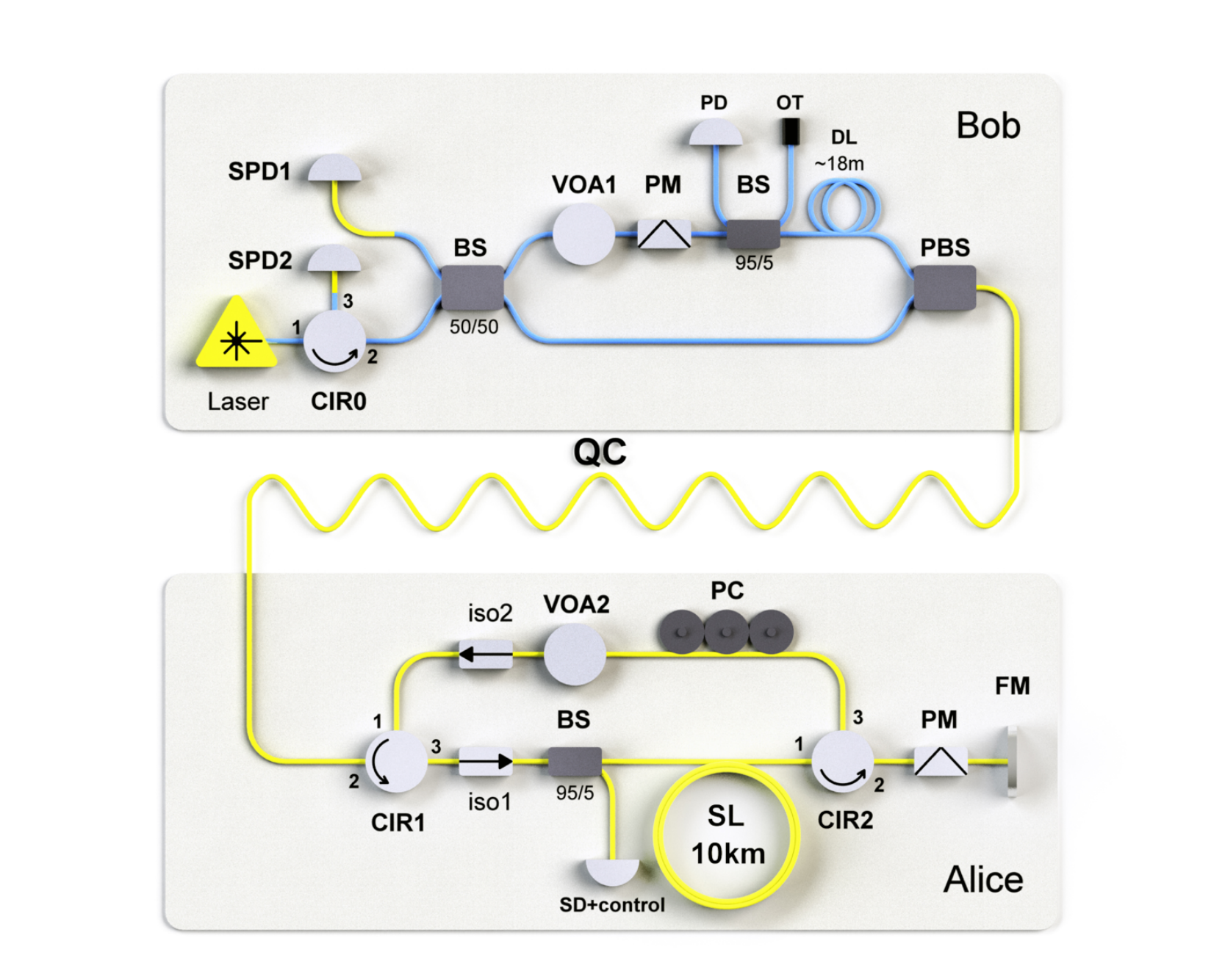}
\put(-340,220){(a)}
\put(-340,65){(b)}
\caption{The fiber-optic go-and-return scheme, consisted of circulators (CIR0, CIR1), single photon detectors (SPD1, SPD2), variable optical attenuators (VOA1, VOA2), phase modulators (PM), beam splitters (BS), photodetector (PD), optical terminator (OT), delay line (DL), polarization beam splitter (PBS), quantum channel (QC), isolators (iso1, iso2), storage line (SL), Faraday mirror (FM), polarization controller (PC), and 1550\,nm telecom laser. The polarization-maintaining and single-mode fibers are represented in blue and yellow colors respectively. For more details and specifications see the main text.}
\label{fig:Setup}
\end{figure}

We took the modular QKD platform for research and education \cite{Rodimin2019-1, Rodimin2019-2} as a basis for our experimental setup. Component specifications not listed here, as well as automation of control and data acquisition, are described in \cite{Rodimin2019-2}. The two-pass auto compensation plug\&play scheme \cite{Stucki2002} was modified to generate both SP and SRP (see Fig.~\ref{fig:Setup}). The attenuator VOA1 (Exfo FVA600) was placed in one of the arms of Bob's interferometer in order to adjust the SP/SRP ratio. Some of the implementation details, in particular the location of VOA1, we reproduced from Ref.~\cite{Legre2006}, where a similar two-pass QKD scheme with two pulses of significantly different intensities was built. Although the scheme in \cite{Legre2006} was  designed for continuous variables QKD with a homodyne detection, replacing the homodyne detector by single photon detectors (SPD) and making several other modifications, discussed below, makes the optical scheme suitable for SR QKD. 

Due to significant difference between the SP and SRP intensities ($t\gtrsim60$\,dB), any reflections of SRP (e.g. on connectors) can have much larger intensity than the signal and thus make the signal undetectable. In order to get rid of such backside reflection and Rayleigh backscattering effects, it is crucial to avoid the overlap between pulses travelling to Alice and those already returning back to Bob. The well-known solution for plug\&play QKD schemes with weak pulses is to send pulses in trains and use a storage line (SL) on Alice's side. However, the relation between SRP and SP is preserved in the storage line as well, keeping the same issue with the parasitic back reflectons, which makes this solution not sufficient enough and hence requires some additional counter-measure.

To solve this issue we added a bypath on Alice's side using two single-mode circulators, denoted in Fig.~\ref{fig:Setup}b as CIR1 and CIR2 (Opto-Link OLCIR-I-3-155). The measured isolation between ports~1 and 3 of the circulators is more than 70\,dB. A train of pulses, sent by Bob, passes through CIR1 and then entirely fits inside the SL. So when the head pulses arrive to the output of CIR1 via the bypath on the way back after being reflected on the Faraday mirror (FM), there will be no overlap with the tail of the train, and all background reflections will be blocked by the isolator iso1.
The allowed number of pulses per train, $N_{\rm p}$, is determined by the SL length: $N_{\rm p}Tc/n_{\rm fib}\lesssim\ell_{\rm SL}$, with optical fiber refractive index $n_{\rm fib}=1.47$ and pulse period $T=f^{-1}$. For $\ell_{\rm SL}=10$\,km and $f=5$\,MHz we obtain $N_{\rm p}\lesssim245$ pulses.
It is important to note that by introducing the bypath we break the polarization auto-compensation of the plug\&play scheme. Therefore, in order to restore the polarization transformations in the SL, a polarization controller (PC) has to be used.

Alice prepares quantum states of qubits using phase coding. For this purpose she randomly chooses 1 of 4 phase shifts $0, \pi, \pi/2, 3\pi/2$ between SRP and SP. This shift is applied on SP using the phase modulator (PM) in the section between CIR2 and Faraday mirror (FM). Bob chooses qubit measurement basis applying with PM 0 or $\pi/2$ phase shift on SRP when it passes through the long arm of Mach-Zehnder interferometer on the way back.

\subsection{Setup adjustment}


The typical value of reflections from FC/PC connectors and some other extinction parts of optical components is about 20\,dB, which is 4-5 orders of magnitude less than the required SRP and SP difference. This must be taken into account when preparing the system. Besides usual for QKD synchronization signals settings, the laser pulses of such a huge difference in the optical scheme lead to that the following preparatory procedures come to the fore:
\begin{itemize}
    \item Optical signals reflections reducing from connectors
    \item SP and SRP pulses isolating in time-scale and by polarization
    \item Detector time-window minimization, particularly, when using a free-run detector
\end{itemize}
Below we describe these refinement procedures.

Because of intensive SRP pulse, many reflections of unclear origin return back to the detectors, shading the signal. One needs to figure out where the parasitic reflections come from that superimpose with SP. Then the reflections may be reduced by cleaning and polishing the connectors, replacing the FC/PC connectors with APC ones, or even fusion splicing of fibers.

In order to understand the origin of various reflected signals, we simulated reflections in our optical scheme \cite{RodiminCode}. The simulation demonstrated that the most critical parasitic reflections originate from the connectors on Alice's side between CIR2 and FM. A replacement of FC/PC by APC connectors in this section provided a sufficient purification of the signal.

\begin{figure}[t!]
\centering\includegraphics[width=0.7\textwidth]{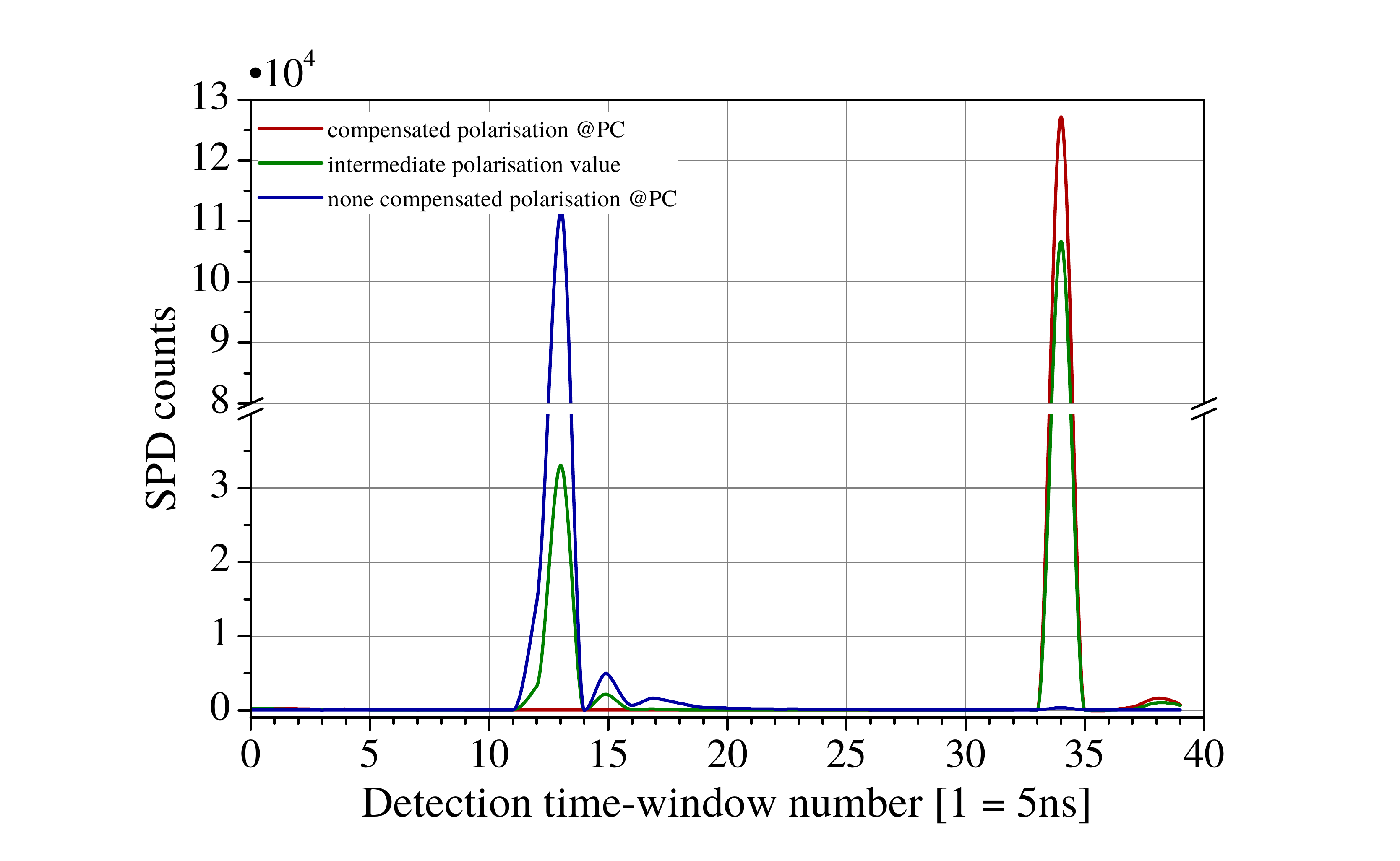}
\caption{Alice polarization controller can be tuned by minimizing the signal measured in an advance time-window, corresponding to difference between the arms of the interferometer. 
}
\label{fig:Tuning}
\end{figure}

When the polarization controller is not adjusted to restore polarization changes, all pulses return from Alice to Bob's PBS in an arbitrary state of polarization. Therefore, SRP has a nonzero projection onto the polarization state which corresponds to the short unattenuated interferometer arm. As a result, some fraction of SRP with corresponding polarization component passes through PBS and arrives at BS via the short arm ahead of schedule, i.e. when there is no pulse coming from the other arm to interfere with. On the reflectogram in Fig.~\ref{fig:Tuning}, obtained by detection window scanning, one can see that the outstripping pulse forms a peak before the signal one, according to the arm length difference (21~m in our case). Thus, the PC state is tuned to minimize the outstripping peak intensity, in this way making the pulses enter the correct arms of the interferometer and as a result providing the maximum QKD rate.

For weak coherent pulses detection in the interferometric scheme, the four-channel free-run superconducting nanowire single-photon detector (SNSPD) Scontel FCOPRS-CCR-TW-60/0.01 was used. To reduce the noise acquisition, each SNSPD click from a laser pulse of width 3\,ns is recorded within a short 5\,ns time window.
Beyond these periodic detection time windows the free-run detector clicks as well, but these clicks are not recorded to memory. For noise reduction, it is important to make the registration slot as short as possible. However, the signal front jitter of the SNSPD we used appeared to be comparable with our 5\,ns window. Therefore, a comparator with essentially lower jitter was constructed and put in one channel of the SNSPD to digitize the analog signal. The key generation results described below are obtained using one SNSPD channel, randomly applying four phase shifts with Bob’s PM; this in turn leads to two-times lower QKD rate.

\begin{figure}[t!]
\centering\includegraphics[width=0.7\textwidth]{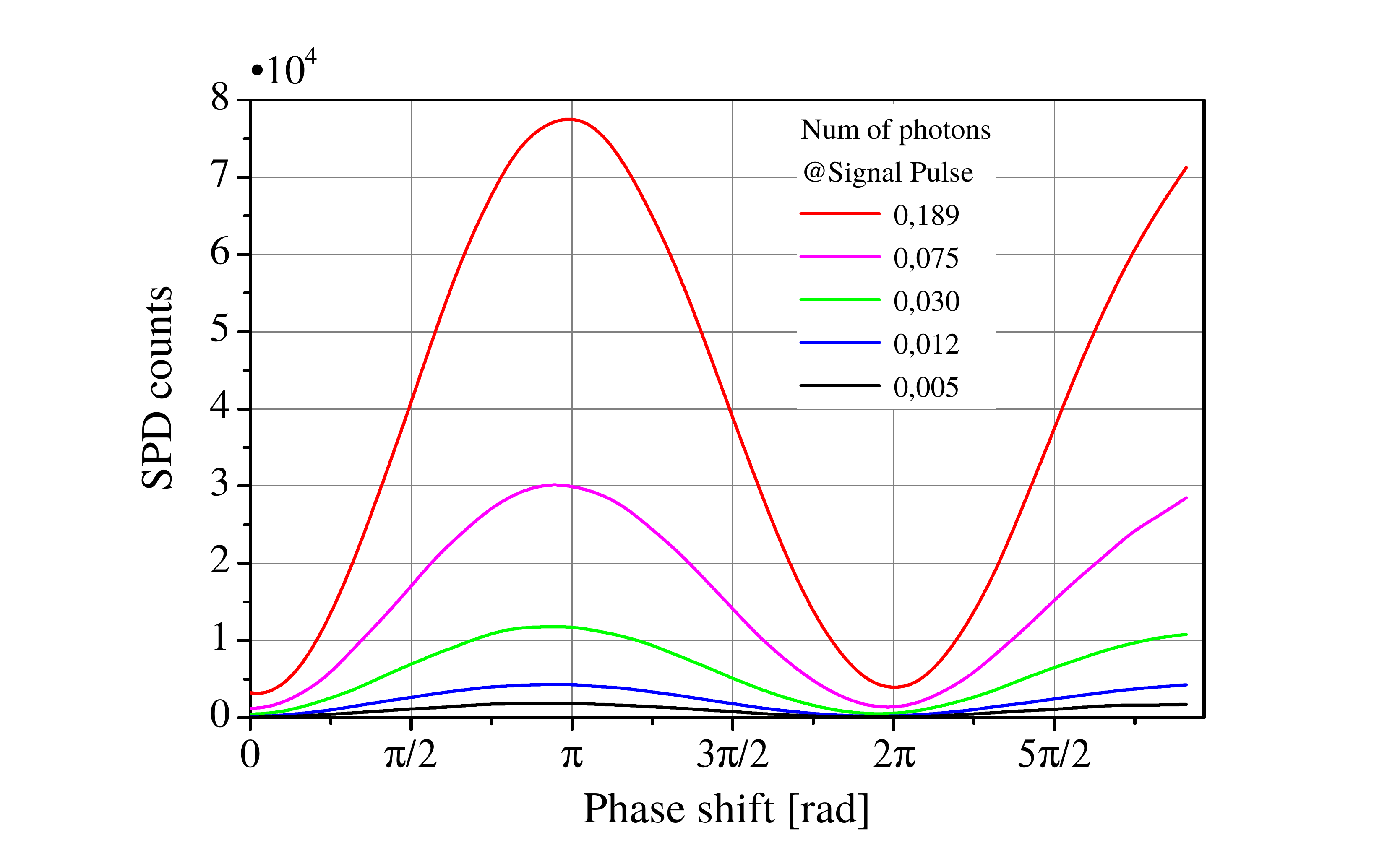}
\caption{Interference curves at various signal brightness: the number of detection clicks as function of the phase shift in the long arm of Bob's interferometer.}
\label{fig:Visibility}
\end{figure}

In order to evaluate optical and noise characteristics of the scheme adjusted with 65\,dB between SP and SRP, we measured the interference curves -- the number of detector counts as function of the SP/SRP phase shift at different laser intensities (see Fig.~\ref{fig:Visibility}). One can see that with increasing the SP intensity not only maximums but also minimums rise as well. This noise increase leads to a bad visibility of the scheme making it unacceptable for QKD. Below we set out the most essential reason leading to this.

\begin{figure}[t!]
\centering\includegraphics[width=0.7\textwidth]{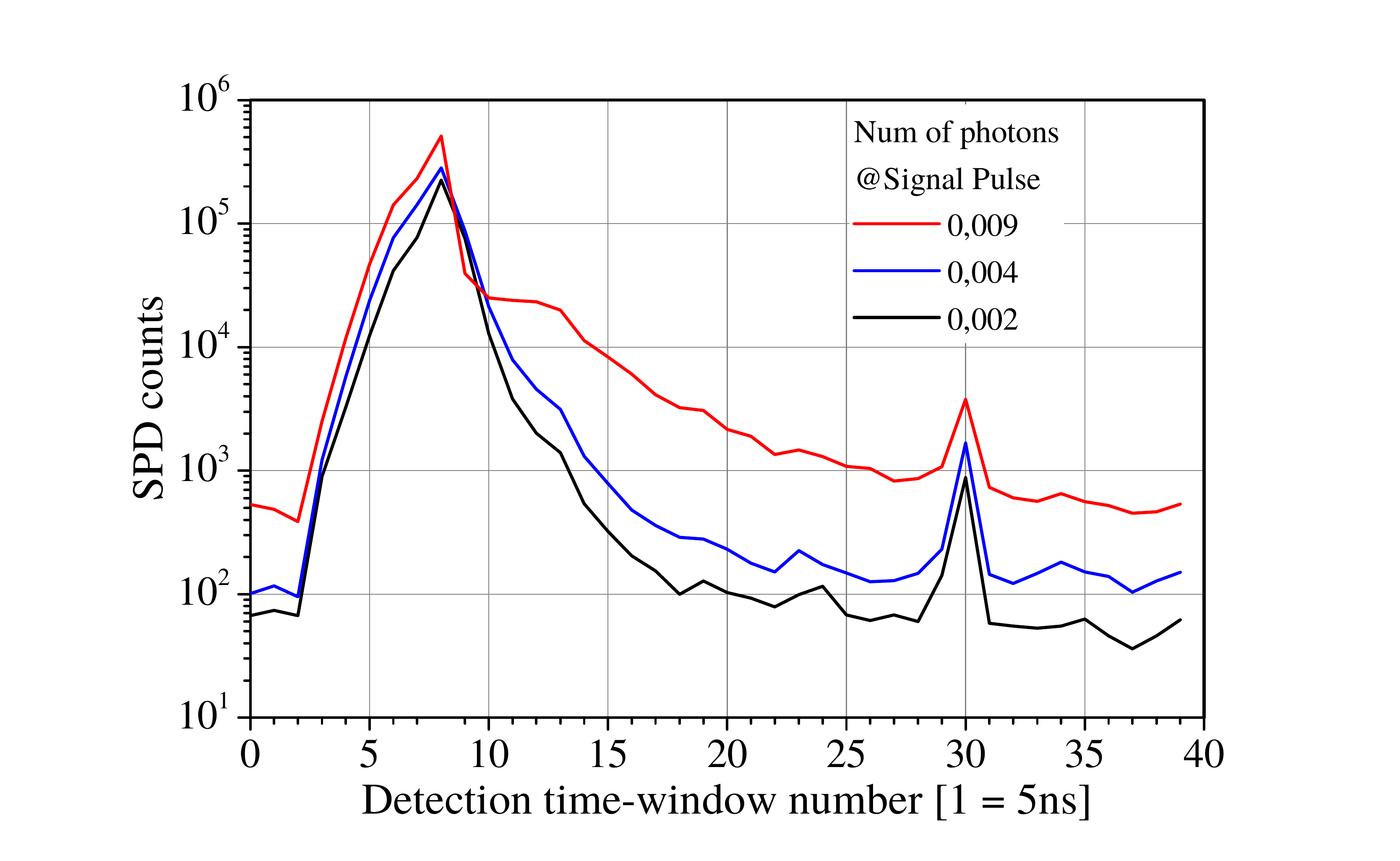}
\caption{SNSPD blinding with the laser pulse brightness increase. A tail of the false clicks is formed after the outstripping parasitic pulse. The signal becomes difficult to distinguish when the tail reaches the signal pulse. Signal scanning was performed within the 200\,ns window with a step of 5\,ns. 
}
\label{fig:Plume}
\end{figure}

Upon photon absorption by a superconducting nanowire of SNSPD, the detector enters a state characterized by formation of a non-superconducting domain. The voltage drop across the nanowire is about 1\,mV. The transition back to the working state is associated with the absorption of heat released on the nanowire and is about several nanoseconds. As the pulse energy increases, it takes more time to absorb this energy and return the detector back to the working state from the metastable one. On the oscillogram, this looks like a noisy tail formation after the front of SNSPD response signal \cite{Elezov2019}. In our case, the presence of outstripping parasitic pulse in 21 meters before the working signal encounters the effect of free-run SNSPD detector blinding. If the outstripping pulse is bright enough, noisy tail after it gives rise to a trail of false detections, which, as the brightness of the pulses increases, reaches the signal pulse, Fig.~\ref{fig:Plume}. With an increase in the SRP/SP ratio it is necessary to adjust the polarization compensation on the PC more and more precisely in order to reduce the outstripping parasitic pulse. But even with an ideal setting, starting from a certain SRP/SP ratio, it is not possible to suppress the outstripping pulse, since the typical value of the PBS extinction is about 23~dB; for SRP/SP 60~dB the intensity of the outstripping pulse is much higher than the signal one. Thus, intensity increasing of the laser pulses leads not only to the stray lighting of the whole line, but due to the detectors blinding, gives rise to the QBER increase. 

\subsection{Scheme operability check results}

\begin{figure}[t!]
\centering\includegraphics[width=0.7\textwidth]{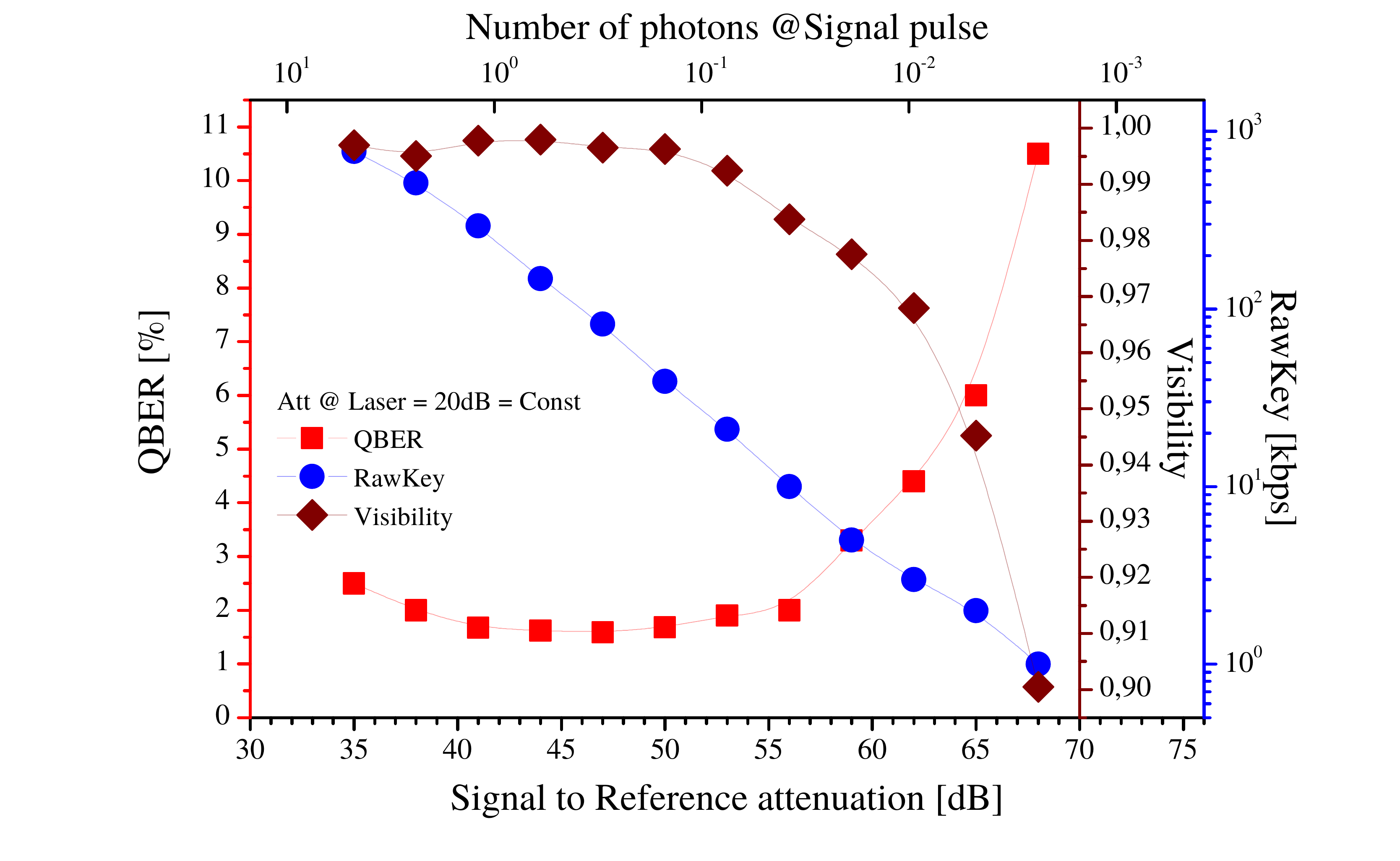}
\caption{Results of experimental key generation in a two-pass scheme using the BB84 protocol with a reference pulse.}
\label{fig:Res}
\end{figure}

For confirming the operability of the scheme, there is no difference which protocol, B92 or BB84 with SRP, to use. We choose BB84 since both the software and the firmware have already been prepared and tested for this protocol. With the plug\&play scheme modification and tuning as described above, the results of the key generation are presented in Fig.~\ref{fig:Res}. We choose the SRP as bright as possible, avoiding the SNSPD blinding. The QBER of 1.5--2\% is characteristic for the contrast of our fiber optic scheme with Mach-Zehnder interferometer. We use the 1\,km long quantum channel. The SNSPD efficiency is set to be 25\% with about 100\,Hz of dark count rate, caused by ambient light as well. As the SP energy decreases, the detector dark counts prevail. Herewith, the QBER starts to grow and the visibility decreases. An acceptable relative difference between SRP and SP of $t=65$\,dB was obtained with $\mu=0.005$ photons per SP. This value would have been increased up to the required optimal level $\mu=\mathcal{O}(0.1)$ if a gated SPD was used. In this way, the blinding effect could be avoided.

\section{Discussion}\label{sec:discussion}

As stated above, for the SR QKD security, the SRP must be reliably detected on Bob's side. In the developed scheme, this is done using a photodetector, denoted as PD in Fig.~\ref{fig:Setup}a. Most of the SRP power goes to the PD when the train comes back. In our case, the accuracy of the SRP control $\delta$ becomes a key part of the security consideration, determining the relative ratio between the SRP and SP intensities $t$ and the maximum QKD distance. In direct photodetection of a weak optical signal, the main noise source is the thermal noise \cite{Hui2019}. To fairly introduce it in our theoretical model, we use the typical NEP value of high-quality photodetectors available on the market.

To deprive Eve of the opportunity to amplify the SP on the way from Bob to Alice, it is necessary to control the SP on Alice's side. This is fraught with technical difficulties; in this work, we have not made a technical implementation of this solution. The problem is that the SP intensity coming to Alice should be high enough for measurement, while the laser power should be enough to generate the SRP. We can assume that Eve does not amplify the SP if we do not distinguish it against the background of the photodiode thermal noise while making the intensity of SP equal to the intensity of the thermal noise. Estimated at a load resistance of $50\,\Omega$ and room temperature, the sensitivity limit of direct photodetection against background of thermal noise corresponds to the laser pulse energy of $10^3$ photons. Then, with $t=60$\,dB and the quantum channel length $L=20$\,km, the peak laser power should be about 150\,mW. This is quite achievable, the DFB Diode Lasers on the market have a peak power of several times more.  We see also other solutions to the SP control problem, namely the use of other powerful lasers, EDFA, multi-pixel avalanche photodetector, SP+SRP homodyning. But these solutions lead to higher costs and/or complications of equipment.

\begin{figure}[t!]\centering
	\includegraphics[width=0.45\textwidth]{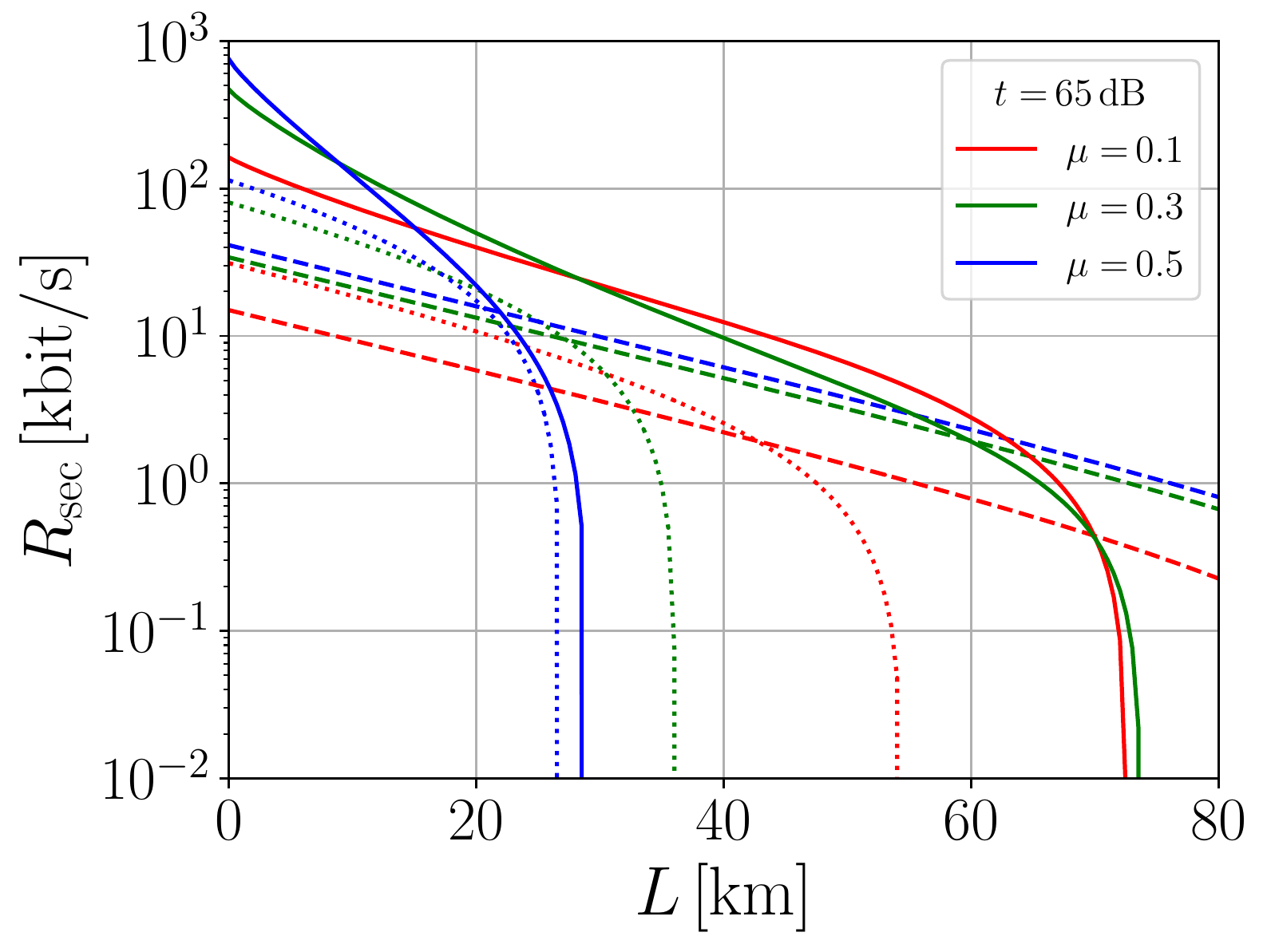}
	\caption{Distance dependence of secure key generation rate. Solid, dotted and dashed lines represent B92 with SRP, standard BB84 and decoy-state BB84 protocols respectively. For the decoy-state protocol we fix $\nu_2:\nu_1:\mu=1:25:100$ and $p_\mu=0.5$.}
	\label{fig:R_sec-L}
\end{figure}

In terms of technical implementation, the realized SR BB84 scheme does not differ much from the B92 protocol with SRP. As for the raw key rates, there is factor two reduction for BB84 due to the basis reconciliation. Therefore, from practical point of view, it seems not reasonable to use the four-state BB84 scheme. In Fig.~\ref{fig:R_sec-L} we compare the secret key generation rates for the SR B92 protocol (given by Eq.~\eqref{eq:R_sec}) and classic BB84 with and without decoy states (see Appendix~\ref{app:BB84}). One can see from the plot that for the short distances up to 60\,km the SR B92 has a higher key rate, making it more advantageous even with respect to the decoy-state method. However, as it was mentioned in the introduction, it lacks the rigorous unconditional security proof. One can also mention that for $\mu=\mathcal{O}(0.1)$ the optical error contribution to QBER (the term proportional to $p_{\rm opt}$ in Eq.~\eqref{eq:QBER}) is practically constant and is dominant for short and mid-range distances, while at long distances the dark count contribution becomes crucial. Therefore, even with perfect single-photon detectors there is a lower bound on QBER due to imperfect visibility of optical scheme. 

\section{Conclusion}\label{sec:conclusion}

Here we have presented the theoretical consideration of SR QKD security based on the properties of realistic photon detectors. For the given setup parameters the optimal photons number values in SP were found to be from 0.15 for 50\,km to 0.35 for 10\,km at 65\,dB between SRP and SP. It was demonstrated that secure key generation is possible only with a bright SRP starting from $4\times10^4$ photons per pulse at 10\,km and more than $5\times10^5$ photons per pulse are needed to maximize the secure key generation rate at 50\,km. Our study shows that SR QKD has advantages for short distances. At 20\,km, the B92 with SR protocol has more than four times higher generation speed compared to the decoy-state BB84 protocol, while at 60\,km, their rates become equal.

We have to emphasize that the above consideration of the security of SR QKD is not a strict unconditional security proof. Our approach relies on particular Eve's attack using the soft filtering operation which is very effective but might be not the ultimate attack on SR QKD. We use the characteristics of realistic equipment available in our laboratory to parametrize the formula for the secret key generation rate. In that sense, the obtained results may be regarded as a somewhat pessimistic estimate, since for theoretical calculations one can use the best attainable characteristics of the equipment.

We have developed the go\&return optical scheme which allows forming laser pulse trains for SR QKD. This scheme was tested by running the sifted key generation, demonstrating the technical feasibility of a 65\,dB difference between SP and SRP. The effect of blinding of the free-run SNSPD we used did not allow us to reach a bigger SRP/SP difference, as well as to generate more than 0.005 photons in SP at 65\,dB. We believe that using of a gated detector will solve this issue. Mostly, the scheme setup is identical to the well-researched and commercially available plug\&play scheme. The frequency of the polarization correction is related to the thermal stability of the segment responsible for the polarization distortion, mainly the storage line. In our experience, the characteristic drift time of polarization, even in field conditions, might be minutes per tens of kilometers.

When developing the optical scheme, one of our motivations was to find and test a practical implementation for CV QKD. We plan to adopt this scheme for homodyning Gaussian-modulated coherent states.

\section*{Acknowledgments}
This work is supported by the Russian Science Foundation under project 17-71-20146.

\section*{Disclosures}
The authors declare no conflicts of interest.

\section*{Data availability}
Data underlying the results presented in this paper are not publicly available at this time but may be obtained from the authors upon reasonable request.

\appendix

\section{Non-orthogonal state discrimination}\label{app:B92}

In the B92 protocol Alice sends two non-orthogonal states, $|\psi_0\rangle$ and $|\psi_1\rangle$, where $|\langle\psi_0|\psi_1\rangle|=\cos\gamma$ ($0<\gamma<\pi/2$) is the main parameter of the protocol. Bob on his side performs a three-outcome measurement, described by a positive-operator-valued measure (POVM) \cite{Peres93} (for the first studies of non-orthogonal state discrimination see \cite{Ivanovic87,Peres88,Dieks88}),
\begin{equation}
    \begin{split}
        M_0 &= {|\psi_1^\perp \rangle\langle \psi_1^\perp| \over 1 + \cos\gamma} = {\mathbb{1} - |\psi_1 \rangle\langle \psi_1| \over 1 + \cos\gamma} \,, \\
        M_1 &= {|\psi_0^\perp \rangle\langle \psi_0^\perp| \over 1 + \cos\gamma} = {\mathbb{1} - |\psi_0 \rangle\langle \psi_0| \over 1 + \cos\gamma} \,, \\
        M_? & = \mathbb{1} - M_0 - M_1 \,,
    \end{split}
    \label{eq:POVM}
\end{equation}
where $M_{0(1)}$ gives a conclusive 0(1)--bit measurement result, while $M_?$ denotes an inconclusive result that does not provide any knowledge about the encoded bit. If Alice sends a state $|\psi_i\rangle$, the corresponding $X$--outcome probability is determined as $p_{X|i}={\rm Tr}(M_X|\psi_i\rangle\langle\psi_i|)=\langle\psi_i|M_X|\psi_i\rangle$. One can easily verify that $\langle\psi_1|M_0|\psi_1\rangle=\langle\psi_0|M_1|\psi_0\rangle=0$, i.e. applying $M_0$ to measure $|\psi_1\rangle$ there is no chance to obtain the outcome ``0'' and vice versa. The inconclusive result probability is given by
\begin{equation}
    p_? = \langle\psi_0|M_?|\psi_0\rangle = \langle\psi_1|M_?|\psi_1\rangle
    = \cos\gamma \,.
\end{equation}
and the probability of a conclusive result is 
\begin{equation}
    p_{\rm ok} = {1\over2} \langle\psi_0|M_0|\psi_0\rangle + {1\over2} \langle\psi_1|M_1|\psi_1\rangle
    = 1 - p_? = 1 - \cos\gamma \,.
\end{equation}

Now let us consider two particular non-orthogonal states, $|\psi_0\rangle=|\alpha\rangle$ and $|\psi_1\rangle=|-\alpha\rangle$ with $|\alpha|^2=\mu$. Writing these coherent states as
\begin{equation}
    |\pm\alpha\rangle = e^{-\mu/2} \sum_{n=0}^\infty {(\pm\alpha)^n \over \sqrt{n!}} |n\rangle \,,
\end{equation}
one finds the overlap between these states,
\begin{equation}
    \cos\gamma = \langle\alpha|-\alpha\rangle = e^{-\mu} \sum_{n,m=0}^\infty {\alpha^{*n} \over \sqrt{n!}} {(-\alpha)^m \over \sqrt{m!}} \langle n|m\rangle = e^{-2\mu} \,.
\end{equation}
Note that the equations above are valid only for the limit case of lossless channel and ideal Bob's detector. So taking into account the optical losses and detector's efficiency, the realistic bit acceptance rate is
\begin{equation}
    p_{\rm ok} = 1 - e^{-2\eta\mu^\prime} \,,
    \label{eq:p_ok_B92}
\end{equation}
where $\mu^\prime$ is the expected signal intensity on Bob's side. Neglecting dark counts, the raw key rate can be estimated as $R_{\rm raw}\simeq fp_{\rm ok}$, where $f$ is the laser pulse repetition frequency.

In the 4+2 protocol \cite{Huttner95} the basic ideas of BB84 and B92 are combined: Alice randomly chooses a basis and a non-orthogonal state in this basis, $\{|\psi_0^z\rangle,|\psi_1^z\rangle\}$ and $\{|\psi_0^x\rangle,|\psi_1^x\rangle\}$, with overlap parameter $|\langle\psi_0^z|\psi_1^z\rangle|=|\langle\psi_0^x|\psi_1^x\rangle|=\cos\gamma$. For instance, one can use laser pulses with $\{0,\pi\}$ and $\{\pi/2,3\pi/2\}$ phase shifts for $z$- and $x$-basis respectively: $|\psi_{0,1}^z\rangle=|\pm\alpha\rangle$, $|\psi_{0,1}^x\rangle=|\pm i\alpha\rangle$. Bob in turn randomly performs one of POVMs to distinguish the states in each set. Due to the basis reconciliation the acceptance rate of conclusive measurements~\eqref{eq:p_ok_B92} has to be multiplied by 1/2 for the 4+2 protocol.

\section{Soft filtering parameter constraints}\label{app:filtering_constraints}


Assuming $|\beta_{s(f)}|^2=\mu_{\max(\min)}^\prime=\mu^\prime(1\pm\delta)$, the {\it success} probability $p$ and the amplification parameter $a$ are determined from Eqs.~\eqref{eq:key_rate_cond} and \eqref{eq:unitarity_cond} respectively:
\begin{equation}
	p = { e^{-2\eta\mu_{\min}} - e^{-2\eta\mu^\prime} \over e^{-2\eta\mu_{\min}} - e^{-2\eta\mu_{\max}} }
	= { 1 \over  1 + e^{-2\eta\mu^\prime\delta} } \,,
\end{equation}
\begin{equation}
	a = -{1 \over 2\mu} \ln \bigg[ {e^{-2\mu} - (1 - p) e^{-2b\mu} \over p} \bigg] = 1 - {\ln \big[ 1 - e^{-2\eta\mu^\prime\delta} \big( e^{2\mu (1 - b)} - 1 \big) \big] \over 2\mu} \,.
	\label{eq:a}
\end{equation}
One can easily verify that for
\begin{equation}
	1 - {\ln \big[ 1 + e^{2\eta\mu^\prime\delta} \big] \over 2\mu} < b < 1 \,,
	\label{eq:b_bound_2}
\end{equation}
the argument of logarithm in Eq.~\eqref{eq:a} is always positive and smaller than 1, and hence $a>1$ as required by the unitarity condition \eqref{eq:unitarity_cond}. Combining the constraint $|\beta_f|<|\alpha_f|$ and Eq.~\eqref{eq:b_bound_2}, one obtains the generalized lower bound,
\begin{equation}
	b > b_{\min} = {\max} \bigg\{ 1 - {\ln \big[ 1 + e^{2\eta\mu^\prime\delta} \big] \over 2\mu} \,, ~(1 - \delta) \times 10^{-0.2L/10} \bigg\} \,.
	\label{eq:b_min}
\end{equation}
For the limit case $|\beta_s|^2=\mu_{\max}^\prime$, one has to take into account the $|\beta_s|^2<a\mu$ constraint as well. As a result, using Eq.~\eqref{eq:a}, one gets the upper bound
\begin{equation}
	b < b_{\max} = \min \bigg\{ 1 - {\ln \big[ 1 - e^{2\eta\mu^\prime\delta} \big( e^{2\mu - 2\mu^\prime (1 + \delta)} - 1 \big) ] \over 2\mu} ,\, 1 \bigg\} \,.
	\label{eq:b_max}
\end{equation}
One can see that if $\mu>\mu^\prime(1+\delta)$, i.e. $L>50\log_{10}(1+\delta)$, the logarithm in \eqref{eq:b_max} turns out to be negative, and hence $b_{\max}=1$ must be used due to unitarity condition.

\section{BB84}\label{app:BB84}

The secret key generation rate is given by GLLP \cite{GLLP} formula, rewritten as \cite{Lo05}
\begin{equation}
	R_{\rm sec} = {1\over2} f \big\{ Q_1 \big[1 - H(E_1)\big] - Q_\mu f_{\rm ec} H(E_\mu) \big\},
	\label{eq:R_sec_GLLP}
\end{equation}
where $Q_1$ and $Q_\mu$ are the single-photon and overall gains with corresponding error rates $E_1$ and $E_\mu$ respectively (for definition see Refs.~\cite{Lo05,Ma05}). In particular, with no eavesdropping $Q_\mu$ and $E_\mu$ can be modelled as
\begin{equation}
	Q_\mu = \sum_{n=0}^\infty Q_n = \sum_{n=0}^\infty {\mu^n \over n!} e^{-\mu} \big[ Y_0 + 1 - \big(1 - \eta \times 10^{-0.2L/10} \big)^n \big] = Y_0 + 1 - e^{-\eta\mu^\prime} \,,
	\label{eq:Q_mu}
\end{equation}
\begin{equation}
	E_\mu = {1 \over Q_\mu} \sum_{n=0}^\infty E_n Q_n = {{1\over2} Y_0 + p_{\rm opt} (1 - e^{-\eta\mu^\prime}) \over Y_0 + 1 - e^{-\eta\mu^\prime}} \,,
	\label{eq:E_mu}
\end{equation}
where $Y_0=2p_{\rm dc}$ is the dark count yield. For simplicity we neglect the effect of afterpulses, stray light and other potential background contributions.

In the standard BB84 without decoy states $Q_1$ and $E_1$ cannot be determined experimentally, only $Q_\mu$ and $E_\mu$ can. Therefore, in order to estimate them we consider the following optimal PNS attack: Eve blocks significant fraction of single-photon pulses and splits all multi-photon pulses, leaving one photon in her quantum memory and forwarding the rest to Bob via lossless channel. To be more realistic, we assume that Eve cannot break in Bob's device and control/replace photon detectors. For this scenario we get the following estimation of the lower bound on $Q_1$ and the upper bound on $E_1$:
\begin{equation}
	Q_1^l = Q_\mu - \sum_{n=2}^\infty {\mu^n \over n!} e^{-\mu} \big[ 1 - (1 - \eta)^{n-1} \big] = Q_\mu - 1 + {e^{-\eta\mu} - \eta e^{-\mu} \over 1 - \eta} \,,
	\label{eq:Q1_PNS}
\end{equation}
\begin{equation}
	 E_1^u = {Q_\mu E_\mu \over Q_1^l} \,, 
	 \label{eq:E1_PNS}
\end{equation}
giving the lower bound on the secret key rate,
\begin{equation}
	R_{\rm sec}^l = {1\over2} f \big\{ Q_1^l [1 - H(E_1^u)] - Q_\mu f_{\rm ec} H(E_\mu) \big\} \,.
	\label{eq:R_sec_l}
\end{equation}

One of the advantages of the decoy-state method is that it allows to determine the lower bounds on background and signal single-photon yields and the upper bound on $E_1$ directly from experiment~\cite{Ma05}: 
\begin{equation}
	Y_0^l = {\max} \bigg\{ {\nu_1 Q_{\nu_2} e^{\nu_2} - \nu_2 Q_{\nu_1} e^{\nu_1} \over \nu_1 - \nu_2}\,, 0 \bigg\} \,,
	\label{eq:Y0_decoy}
\end{equation}
\begin{equation}
	Q_1^l = {\mu^2 \, e^{-\mu} \over (\nu_1 - \nu_2) (\mu - \nu_1 - \nu_2)} \bigg[ Q_{\nu_1} e^{\nu_1} - Q_{\nu_2} e^{\nu_2} - {\nu_1^2 - \nu_2^2 \over \mu^2} (Q_\mu e^\mu - Y_0^l) \bigg] \,,
	\label{eq:Q1_decoy}
\end{equation}
\begin{equation}
	E_1^u = { (E_{\nu_1} Q_{\nu_1} e^{\nu_1} - E_{\nu_2} Q_{\nu_2} e^{\nu_2}) \mu\, e^{-\mu} \over (\nu_1 - \nu_2) Q_1^l} \,,
	\label{eq:E1_decoy}
\end{equation}
where $\mu$, $\nu_1$ and $\nu_2$ are the signal, weak decoy and vacuum decoy state intensities respectively. Since the state type is chosen randomly with corresponding probability, the key rate \eqref{eq:R_sec_l} must be multiplied by additional factor $p_\mu$. For illustration, we fix $\nu_2:\nu_1:\mu=1:25:100$ and $p_\mu=2p_{\nu_{1,2}}=0.5$. Further parameter optimization and taking into account statistical fluctuations are beyond this work.

\bibliography{PlugAndPlay_SR_QKD}

\end{document}